\documentclass[12pt]{article}

\usepackage{graphicx,psfrag,epsf}
\usepackage{endfloat}
\usepackage{amssymb}
\usepackage{subfig}
\usepackage{float}
\usepackage{caption}
\usepackage{mathtools}
\usepackage{amsmath,esint}
\usepackage{array}
\newcolumntype{P}[1]{>{\centering\arraybackslash}p{#1}}
\newcolumntype{M}[1]{>{\centering\arraybackslash}m{#1}}

\usepackage[default]{jasa_harvard}    
\usepackage{JASA_manu}

\begin{document}

\title{Double-Crossing Benford's Law}
\author{Javad Kazemitabar  
\thanks{
       Javad Kazemitabar is with the EECS department at Babol Noshirvani University of Technology, Babol, Iran. 
     \texttt{j.kazemitabar@nit.ac.ir}.}
}

\maketitle


\mbox{}

\begin{abstract}
Benford's law is widely used for fraud-detection nowadays. The underlying assumption for using the law is that a "regular" dataset follows the significant digit phenomenon. In this paper, we address the scenario where a shrewd fraudster manipulates a list of numbers in such a way that still complies with Benford's law. We develop a general family of distributions that provides several degrees of freedom to such a fraudster such as minimum, maximum, mean and size of the manipulated dataset. The conclusion further corroborates the idea that Benford's law should be used with utmost discretion as a means for fraud detection.
\end{abstract}
$\bf{Keywords:}~$Benford's law, fraud detection, statistical analysis
\noindent%

Ever since Varian \cite{HalVarian} suggested using Benford's law for detecting economic fraud, researchers have explored many other areas to apply the law. Examples include -but by no means limited to- checking national Covid-19 mortality rate \cite{NatureCovid}, tax \cite{NigriniTax}, and financial statements fraud\cite{Financial}. Benford's law has also been used to investigate Natural Hazard dataset homogeneity \cite{NaturalHazard}. Deviation of a list of numbers from Benford's law is usually considered as a red flag. However, it has been suggested \cite{russianElection} that perfect adherence to this law could also imply manipulation as we expect small levels of deviation from the law in regular lists of numbers. A question is then raised as whether it is possible to systematically manipulate a list such that it still complies with the law. In this paper we address this question and show that it is possible to deceive an auditor by generating a Benford-compliant list with desired statistics, such as max, min, average and size. We do so by building a distribution with tunable parameters that provide such degrees of freedom. 
\section*{Benford-compliant distributions}
Hill's 1995 paper \cite{HillPaperStat95} provides a statistical explanation of Benford's law. The author shows that ``if probability distributions are selected at random, and random samples are then taken from each of these distributions in any way so that the overall process is scale (or base) neutral'' then Benford's law holds. He then asks ``An interesting open problem is to determine which common distributions (or mixtures thereof) satisfy Benford's law''. Several researchers pursued this question and found conditions for a Benford-compliant distribution \cite{Balanzario2010}\cite{Balanzario2015}\cite{commontwo12}\cite{HillFourier}. They also proposed example distributions that satisfy these conditions. However, the proposed distributions do not provide the necessary degrees of freedom for a fraudster to build synthetic \footnote{The term \it{synthetic} Benford set was first used by the celebrated author Mark Nigrini \cite{NigriniBook11}. He provides a method based on the uniform mantissa concept to build synthetic Benford-compliant samples, where the user can designate the maximum and minimum of the generated numbers.} Benford-compliant samples with desired statistics. In this paper, we provide two families of Benford-complaint distributions with tunable parameters. The mere existence of such distributions shows that Benford's law should be carefully used as a means of fraud detection.

\subsection*{Building upon existing proposed distributions}

In \cite{commontwo12}, a few Benford-compliant distributions were proposed that are the building blocks of the distributions to be introduced in this paper.

\begin{itemize}
\item \textit{Example 1:} Let $Y \sim U(0,2)$. Then, $X = 10^Y$ is a Benford compliant distribution defined in $(10^0, 10^2)$. The result can be generalized for $Y \sim U(a,b)$ for integer $a$ and $b$.
\item \textit{Example 2:} Let $Y \sim Triangular(0,1,2)$  In other words:
\begin{equation}
     f_Y(y) =  \left\{
     \begin{array}{ll}
 y & 0 < y < 1 \\
 2-y & 1 \leq y < 2
 \end{array}
 \right. 
 \label{Example2}
 \end{equation}
 Then, $X = 10^Y$ is a Benford compliant distribution defined in $(10^0, 10^2)$. The result can be generalized to symmetric Triangular distributions of $Y$ such as $Triangular(a,b,c)$ where $a$, $b$, and $c$ are all integers and $b=(a+c)/2$.
\end{itemize}

In both of the above examples, even though the maximum and minimum of the distribution -in its general form- is tunable, the average is not. To amend this shortcoming we use the a lemma that was independently proven by a number of authors \cite{OurPaper}\cite{Balanzario2010}\cite{Balanzario2015}.

\textit{Lemma:} If 
\begin{equation}
    \sum_{k=-\infty}^{+\infty}{f_Y(z+k)} =  1
\end{equation}
Then, $X = 10^Y$ is a Benford compliant distribution. Using this lemma, we build upon these examples to introduce our tunable distributions. Concretely, we design the distributions such that the shifted versions of the density function add up to 1.

\begin{itemize}
\item Let
\begin{equation}
    f_{Y_1}(y) =  \left\{
    \begin{array}{ll}
\frac{a}{\sum{a^t}} & m \leq y < m+1 \\
\frac{a^2}{\sum{a^t}} & m+1 \leq y < m+2 \\
& \vdots \\
\frac{a^K}{\sum{a^t}}  & m+K-1 \leq y < m+K \\
\end{array}
\right. 
\label{general1}
\end{equation}
Then, $X_1 = 10^{Y_1}$ is a Benford compliant distribution with the following statistics:
\begin{equation}
\begin{array}{l}
    min(X) = 10^m \\
    max(X) = 10^{m+K} \\
    mean(X) = \frac{9a}{Ln(10).\sum{a^t}}.10^m.\frac{1 -(10a)^K }{1-10a}
\end{array}
\end{equation}
where mean($X_1$) ranges between $3.9\times10^m$ and $3.9\times10^{m+K-1}$ for very small and very large values of $a$ respectively. 

\begin{figure}[ht]
\centering
\includegraphics[height=2.8in]{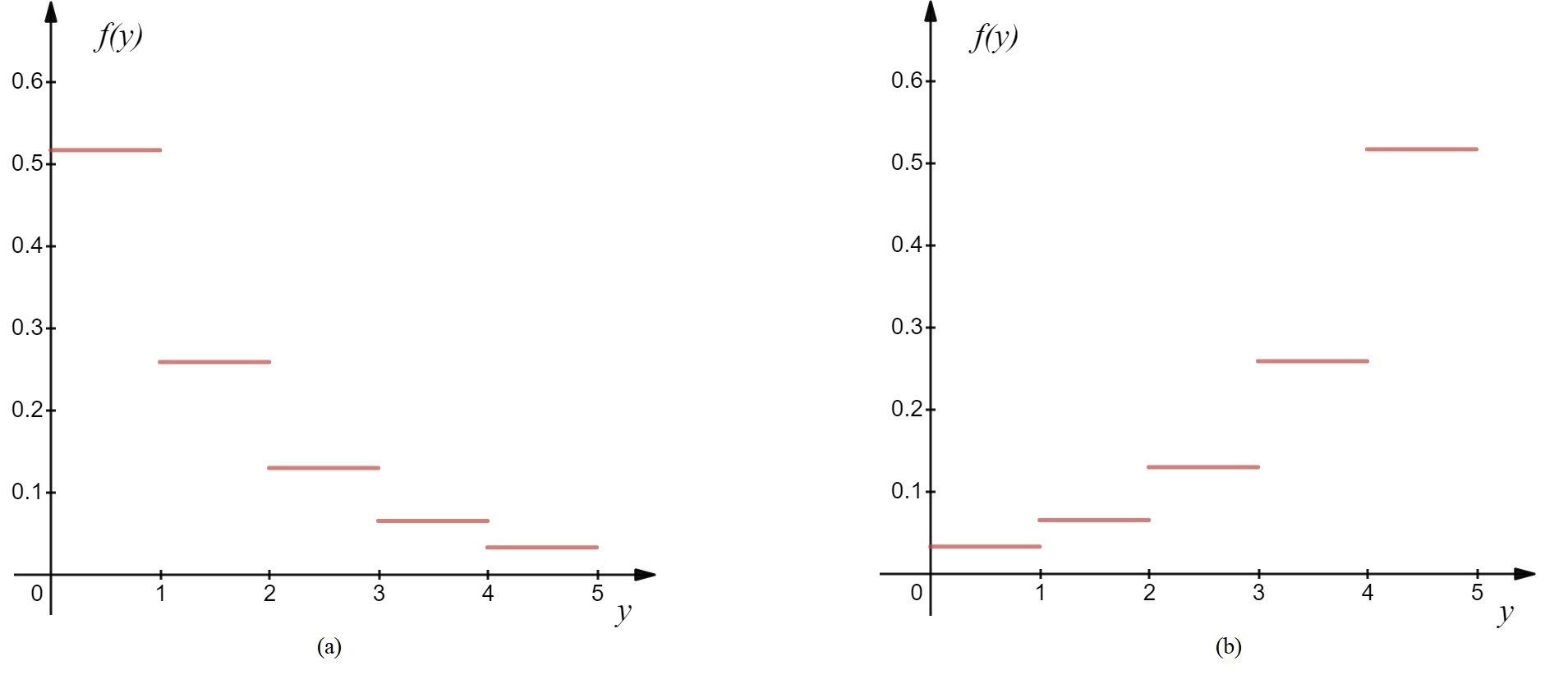}
\caption{An example $Y_1$ distribution with $m=0$, $k=5$ and (a) $a=0.5$ (b) $a=2$. $X_1 = 10^{Y_1}$ follows Benford's law.}
\label{fig:stream}
\end{figure}

\item Let
\begin{equation}
    f_{Y_2}(y) =  \left\{
    \begin{array}{ll}
\frac{a}{\sum{a^t}}(y-m) & m \leq y < m+1 \\
\frac{a}{\sum{a^t}} - \frac{a}{\sum{a^t}}(y - m - 1) & m+1 \leq y < m+2 \\
\frac{a^2}{\sum{a^t}}(y-m-2) & m+2 \leq y < m+3 \\
\frac{a^2}{\sum{a^t}} - \frac{a}{\sum{a^t}}(y-m-3) & m+3 \leq y < m+4 \\
& \vdots \\
\frac{a^K}{\sum{a^t}}(y-m-2K+2) & m+2K-2 \leq y < m+2K-1 \\
\frac{a^K}{\sum{a^t}} - \frac{a}{\sum{a^t}}(y-m-2K+1) & m+2K-1 \leq y < m+2K \\
\end{array}
\right. 
\label{general1}
\end{equation}
Then, $X_2 = 10^{Y_2}$ is a Benford compliant distribution with the following statistics:
\begin{equation}
\begin{array}{l}
    min(X_2) = 10^m \\
    max(X_2) = 10^{m+2K} \\
    mean(X_2) = \frac{99 - 81/Ln(10)}{Ln(10).\sum{a^t}}.10^m.a.\frac{1-(100a)^K}{1 - 100a}
\end{array}
\end{equation}
\end{itemize}
where mean($X_2$) ranges between $2.7\times10^{m+1}$ and $2.7\times10^{m+2K-1}$ for very small and very large values of $a$ respectively.
\begin{figure}[ht]
\centering
\includegraphics[height=2.8in]{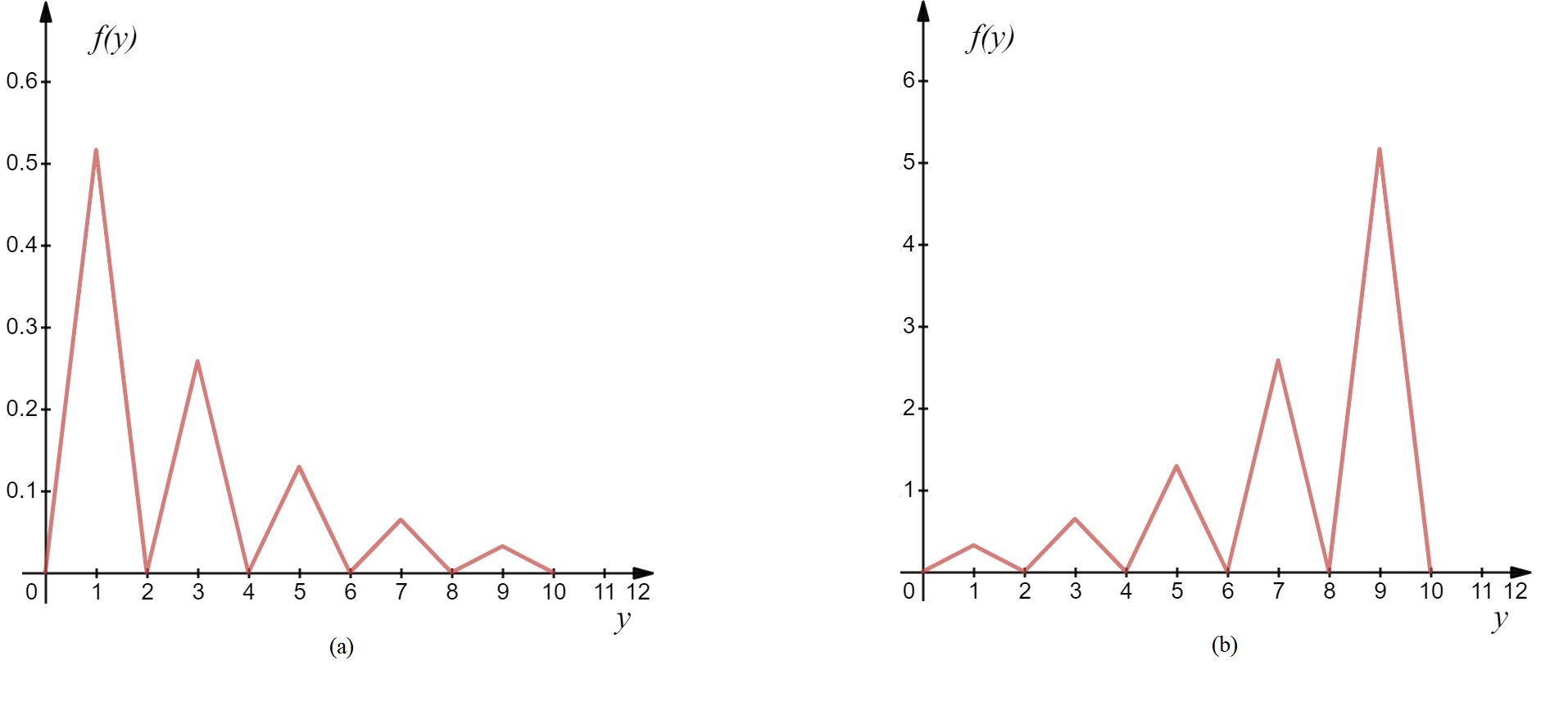}
\caption{An example $Y_2$ distribution with $m=0$, $k=5$ and (a) $a=0.5$ (b) $a=2$.  $X_2 = 10^{Y_2}$ follows Benford's law}
\label{fig:stream}
\end{figure}

One might wonder if the maximum and minimum points in the above mentioned distributions have to be powers of 10. To answer this, we should recall that Benford compliance is scale-invariant. As such the generated numbers can be multiplied with a constant number. Nevertheless, both the above proposed distributions require that the max and min are apart by an integer power of 10, that is $\frac{max}{min} = 10^K$ in the first distribution and $\frac{max}{min} = 10^{2K}$ in the second.
\section*{Building Fake Data}
In this section, we show how a fraudster can generate a Benford-compliant dataset. The generated data could be faked as journal entries of a company trying to look profitable. Of course, in order to fake a journal entry, the fraudster needs to generate two separate datasets; one for income and the other for expenses. For each dataset We can tune the maximum and minimum as well as the number of items and the total sum. This is directly achieved by plugging the right value for $m$, $K$ and $a$ in the distributions introduced in the previous section. Moreover, we note that total sum of numbers in the dataset is equal to the size of that dataset multiplied by its average. Since, we have control over size and average, as a result we have control over total sum. We use Inverse transform sampling \cite{sampling} to generate random samples. In this technique a uniformly generated set of samples is fed into $F^{-1}(y)$, where $F(y)$ is the cumulative distribution function of interest. Now, suppose the hypothetical company's income and expenses each total 5700000\$ and 2310000\$ respectively. Also, let us assume there are 1320 income entries in the journal ranging from 1000\$ to 100000\$ and 760 expense related entries in the range of 100\$ to 100000\$. Using Equation (4), we find $m$, $K$ and $a$ to be 3, 2, and 0.01177886831 respectively for income related entries. Moreover, for expense related entries, we find the aforementioned parameters to be 2, 3, and 0.25927727232382797. While for the scenario at hand we were able to analytically solve for $a$, in general, however, numerical methods may be necessary specially when $K$ is a large number. Figure 3 shows the histograms of income and expense entries. We then generate $X=10^Y$ to populate the journal entries for revenue and expense separately. The total sum for revenue samples add up to 5556356 which is 97$\%$ accurate compared to the requested revenue of 5700000\$. As for the expense dataset, the sum of fake journal entries is 2192381 which shows 5$\%$ deviation from the desired expense total of 2310000\$.  

\begin{figure}[ht]
\centering
\includegraphics[height=2.6in]{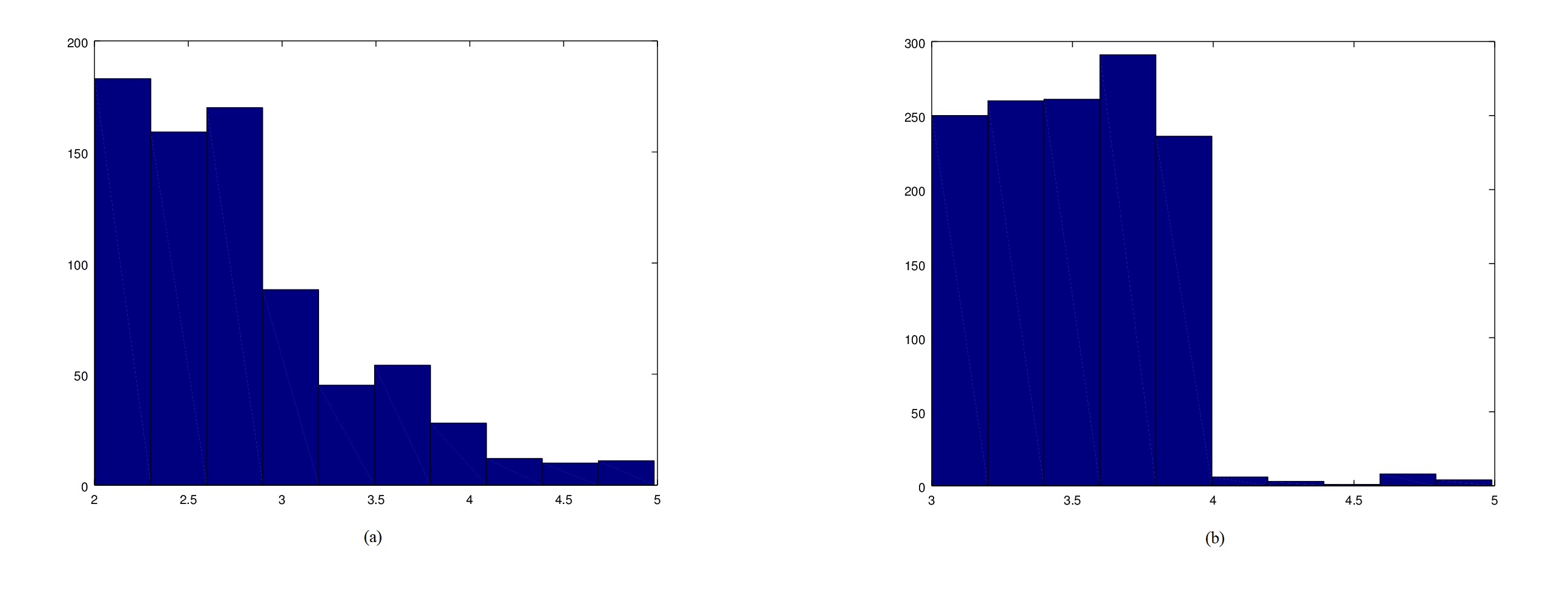}
\caption{Histogram of the synthetic (fake) data generated based on the proposed $Y_1$ distribution. The actual journal entries will be populated by taking 10 to the power of these numbers. (a) Expense related $Y$ samples with $m=2$ and $K=3$ (b) Income related $Y$ samples with $m=3$ and $K=2$.}
\label{fig:stream}
\end{figure}

\section*{Discussion}
We tested the generated journal entries across 3 popular Benford tests namely chi-square, mantissa-arc and mean absolute deviation (MAD). The results of all three tests are shown in Table 1. As can be seen, the generated datasets conveniently pass the Benford test in all the three methods. The practice of generating fake Benford-compliant datasets can easily be performed so long as the average is not too close to either end, i.e. minimum or maximum of the desired set. Concretely, the first proposed distribution, $X_1$, ranges between 3.9 times the minimum value, i.e. $10^m$, and $0.39$ of the maximum value, i.e. $10^{m+K}$. In practical scenarios, it rarely happens that the dataset is skewed to the level that the average exceeds the aforementioned limits. As such, building fake data to deceive the auditor is usually achievable and thus the auditor shall not solely rely on Benford test.  
\begin{table}[ht]
\centering
\begin{tabular}{|l|l|l|l|}
\hline
 & Chi-squared test p-value & Mantissa arc test p-value & Mean Absolute Deviation (MAD) \\
\hline
Revenue & 0.9 & 0.93 & Close conformity \\
\hline
Expense & 0.54 & 0.28 & Acceptable conformity \\
\hline
\end{tabular}
\caption{\label{tab:example}Benford test results confirm compliance of fake data.}
\end{table}

\bibliographystyle{ECA_jasa}
\bibliography{referrences}

\end{document}